# PODIUM: A PULSAR NAVIGATION UNIT FOR SCIENCE MISSIONS

**Francesco Cacciatore[a]\*, Víctor Gómez Ruiz[b], Gonzalo Taubmann[c], Jacinto Muñoz[d], Pablo Hermosín[e], Marcello Sciarra[f], Martiño Saco[g], Nanda Rea[h], Margarita Hernanz[i], Emilie Parent[j], J. Vandersteen[k]**

[a] *SENER Aeroespacial, francesco.cacciatore@aeroespacial.sener*
[b] *SENER Aeroespacial, victor.ruiz@aeroespacial.sener*
[c] *SENER Aeroespacial , gonzalo.taubmann@sener.es*
[d] *SENER Aeroespacial, jacinto.msanchez@aeroespacial.sener*
[e] *Deimos Space SLU, pablo.hermosin@deimos-space.com*
[f] *Deimos Space SLU, marcello.sciarra@deimos-space.com*
[g] *Deimos Space SLU, martino.saco@deimos-space.com*
[h] *Institut d'Estudis Espacials de Catalunya (IEEC) - Consejo Superior de Investigaciones Científicas (CSIC), rea@ice.csic.es*
[i] *Institut d'Estudis Espacials de Catalunya (IEEC) - Consejo Superior de Investigaciones Científicas (CSIC), hernanz@ice.csic.es*
[j] *Institut d'Estudis Espacials de Catalunya (IEEC) - Consejo Superior de Investigaciones Científicas (CSIC), parent@ice.csic.es*
k ESA- ESTEC GNC, AOCS and Pointing division, *jeroen.vandersteen@esa.int*
\* Corresponding Author

**Abstract**

PODIUM is a compact spacecraft navigation unit, currently being designed to provide interplanetary missions with autonomous position and velocity estimations. The unit will make use of Pulsar X-ray observations to measure the distance and distance rate from the host spacecraft to the Solar System Barycenter. Such measurements will then be used by the onboard orbit determination function to estimate the complete orbital elements of the spacecraft. The design aims at 6 kg of mass and 20 W of power, in a volume of 150 mm by 240 mm by 600 mm. The Pulsar X-Ray navigation has been theoretically addressed in various papers and was demonstrated by SEXTANT/NICER on the International Space Station (ISS). The aim of the activity carried out by the SENER-DEIMOS-IEEC industrial consortium under ESA contract is to define a preliminary design for the unit, tackling the overall unit architecture, the optical and thermomechanical design, the unit avionics and SW, and a preliminary concept of function, performance, and operation. PODIUM is designed to minimize the impact on the mission operational and accommodation constraints. The architecture is based on a grazing incidence X-ray telescope with focal distance limited to 50 cm. The effective area shall be in the range 25 to 50 cm2 for photon energies in the range 0.2-10 keV, requiring nesting of several mirrors in the Wolter-1 geometry. Grazing incidence angles will be very small, below 2 deg. The FOV size will determine the aperture of the optics and be related to the maximum number of nested mirrors to accommodate. The current target FOV is 0.25 deg. The pulsars' photon arrivals are detected with a single pixel Silicon Drift Detector (SDD) sensor with timing accuracy below 1μsec. This leads to an expected position accuracy of 30 m. The unit has no gimbaling to meet the applicable power, size and mass requirements. Instead, the host spacecraft shall slew and point to allow pulsar observation. The avionics architecture is based on a radiation hardened LEON4 processor, to allow a synchronous propagation task and measurement generation and orbit determination step in an asynchronous task. PODIUM will enable higher autonomy and lower cost for interplanetary missions. L2 space observatories and planetary flybys are the current reference use cases. Onboard autonomous state estimation can reduce the ground support effort required for navigation and orbit correction/maintenance computation, and reduce the turnaround time, thus enabling more accurate maneuvers, reducing the orbit maintenance mass budget. **Keywords:** Pulsar, navigation, XNAV.

**Acronyms/Abbreviations**

Attitude and Orbit Control System AOCS
Carbon Fiber Reinforced Plastic CFRP
Counts Per Second CPS
Electromagnetic Compatibility EMC
Failure Detection, Isolation and Recovery FDIR
Field Of View FOV
Field Programmable Gate Array FPGA
Finite Element Model FEM
Front End Electronics FEE
Gravitational Wave GW
Homogeneous Poisson Process HPP
Institut d'Estudis Espacials de Catalunya IEEC
International Space Station ISS
Jet Propulsion Laboratory JPL
Line of Sight LoS
Millisecond Pulsars MSPs
Monte-Carlo analysis MC
Multi-Layer Isolation MLI
Neutron Star Interior Composition Explorer NICER
Non-Homogeneous Poisson Process NHPP
On Board Computer OBC
Orbit Determination OD
Phase of Arrivals POAs
Position, Velocity, and Time PVT
Pulsar PSR
Pulsar Timing Array PTA
Random Walk Frequency Noise RWFN
Real-Time Operating System RTOS
Signal to Noise Ratio SNR
Single Layer Isolation SLI
Silicon Drift Detector SDD
Size, Weight, and Power SWaP
Solar System Barycenter SSB
Spacecraft S/C
System On-Chip SoC
Telecommand TC
Telemetry TM
Time of Arrivals TOAs
White Frequency Noise WFN
X-ray Pulsar Based Navigation XNAV

1. **Introduction**

Autonomous celestial-based navigation is considered to be a great alternative to complementing existing orbit determination systems and also for the development of future navigation techniques that allow reducing the dependency from Earth. This would help not only to increase the state knowledge and autonomy of the spacecraft (S/C) in deep space, but also to increase the autonomy level in critical manoeuvres and tasks that require a precise orbit determination solution.

The concept of using pulsars for spacecraft navigation has been in development since the 1960's. Initial studies were conducted by JPL in the 70's and 80's based in both radio and X-ray bands of the electromagnetic spectrum. In case of radio or optical observations from a pulsar, the limitations are very constraining since scattering, dispersion and absorption are rather severe in these energy bands, and the pulsars are relatively faint objects in the Galaxy compared with other celestial objects, increasing the confusion problems for an instrument with small imaging capabilities. Furthermore, the required hardware to detect pulsars in the radio and optical bands would be extremely large to be feasible for flying in space. On the other hand, using X-ray observations, a larger SNR can be achieved for pulsars using much smaller and lighter equipment.

Different algorithms and methods have been explored in order to include the pulsar observations in the orbit determination process. A key step in XNAV is the estimation of the pulse phase, using initial position estimate of the S/C and various ephemeris parameters.

The effect of ephemerides errors, satellite clock-errors, and the movement of the spacecraft during the filtering process can degrade the orbit determination solution accuracy. To eliminate the effect of the S/C motion, the phase estimation can be coupled with Doppler frequency. The Doppler shift can then be converted to speed along the line of sight to the pulsar and thus, having different measurements from different sources will provide observability in all three directions. Notice that in order to obtain a sufficiently precise (low-noise) Doppler measurement, a sufficient number of photons shall be collected, meaning that the duration and planning of the observation campaign will play a relevant role.

If the number of observations is enough, at least four, pulsar navigation can also be used to correct the on-board clock to meet the clock requirements for tracking communication signals. Using pulsar time of arrival and an internal model of the clock, a filtering process can provide the values of the different coefficients of the clock model using the offset between the estimated clock error and the computed clock error as measurements.

Some actual technology demonstrators with flight heritage include NICER/SEXTANT [1][2], the XPNAV-1 [3] and the Insight-HXMT [4] missions. A concept mission with a similar size to the presented in this paper is Cube-X [5].

According to existing experiments of XNAV technology, the measurement noises from pulsar observations cover a big range from several hundred meters to some kilometres depending on the observation times. This leads to orbit determination solutions with accuracy in the order of the km. A summary is provided in Table 1.

Table 1 Summary of performances of XNAV experiments/studies

| Reference | Measurement accuracy | Navigation accuracy 1-σ |
|---|---|---|
| SEXTANT [2] | ~3 km | ~1-5 km |
| [6] | ~0.1-10 km | ~1-10 km |
| [7] | ~1-50 km | ~30-45 km |
| [8] | ~0.8-30 km | ~1.5-5 km |
| Cube-X [5] | ~2-3 km | < 20 km |

No instruments with high Size, Weight and Power (SWaP) constraints have been realized so far.

In this paper, first the unit requirements are presented in Section 2. The high-level concept of the system is explained in Section 3. Then, the preliminary design of the unit is described in Section 4. In Section 5, the Pulsar Sky Catalogue is presented. Finally, the results on the performance of the system are evaluated in Section 6.

The paper ends with the conclusions in Section 7.

## 2. Unit requirements

The objective of this study is to design and evaluate the performance of an autonomous XNAV unit with a 60x15x24 cm$^3$, 6 kg and 20 W limit, able to be integrated in a wide range of satellites.

The design must be assessed in two operational cases: a typical L2 observatory orbit, and planetary fly-by conditions.

The initial accuracy requirements are 10 km position knowledge error with 99% probability at 90% confidence level, at any time during L2 mission nominal science phase or 6 hours before a planetary flyby pericenter.

## 3. High-Level Concept

PODIUM is based on a small telescope with an X-ray detector placed at its focal point. The photons incoming from the pulsar are read at the detector and the timestamps assigned are used by the on-board computer (OBC) to regenerate the signal. This is done over several hours of observation, where the photons are time-folded into a single period until sufficient signal to noise ratio (SNR) is achieved.

Once a clear signal is obtained, it is compared to the reference signal preloaded on the OBC at the observation time. The time shift between these two signals, along with an estimation of the current position of the spacecraft with sufficient accuracy, allows the generation of a ranging measurement that is then used to improve the estimation of the state of the S/C.

With the objective of designing a small unit with high SWaP constraints, and considering similar instruments, the effective area of the telescope will be around 50 cm2. For the same reason, the unit has no gimbaling capabilities; the host spacecraft must be in charge of the pointing maneuvers. In the case some parameters need to be updated from ground on the OBC, all communications will be carried with the host spacecraft and the information will be passed to the unit through a communications link.

A simulator has been developed to assess the performance of the system. The photon sequence that would arrive to the telescope is first simulated taking into account the characteristics of the source Pulsar and those of the unit (mainly the effective area and the selected clock and detector errors). Then, a differentiated onboard algorithm, which does not share variables with the photon sequence algorithm and only uses the timestamps as an input, reconstructs the signal and calculates the phase shift and finally the distance between the spacecraft and the reference position (usually the Solar System Barycenter (SSB)) in the direction of the pulsar. Finally, a separate algorithm performs the orbit determination using the system measurements.

## 4. Preliminary Design

In this section, the preliminary design of the unit is presented. The modes architecture of the system, along with the user operation and system autonomy concept are explained. Then, the measurement and orbit determination algorithms are reviewed. Finally, the opto-mechanics and avionics of PODIUM are presented.

### 4.1. Unit Modes

The operating modes defined for PODIUM are the Observation Mode, the Propagation Mode, the IDLE/Safe Mode, the Calibration Mode, and the Ground Testing Mode.

The Observation Mode is the main operating mode of the unit. In this mode, the telescope is pointed by the spacecraft to a pulsar and the detector is active. The detector reads the incoming photons and the front-end electronics time-tag them. The received photons are time-folded, the signal reconstructed, and the distance to the reference position calculated. During this mode the propagation function keeps providing the state of the spacecraft, being updated when a new measurement is available.

In the Propagation Mode, the system continuously outputs the state (position and velocity) of the spacecraft, calculated by the Orbit Determination function. The base for this propagation is the last update from the observation mode. The spacecraft doesn't need to point the system towards a pulsar and only the OBC must be powered.

The IDLE/Safe Mode is the central mode of the unit, from which every other mode is accessed (via telecommand) and to with the system transitions if a failure occurs (safe mode), or at the completion of a calibration. In this mode, the detector and the front-end electronics can be powered off, and the onboard parameters can be updated via telecommand.

While in the Calibration Mode, a calibration of the detector is performed. Upon finalization, the system transmissions automatically to IDLE/Safe Mode.

Finally, the Ground Testing Mode has the purpose of performing tests to the unit and is only accessible from ground via a special connector plus a telecommand.

Fig. 1 present how is the transition between the different operating modes.

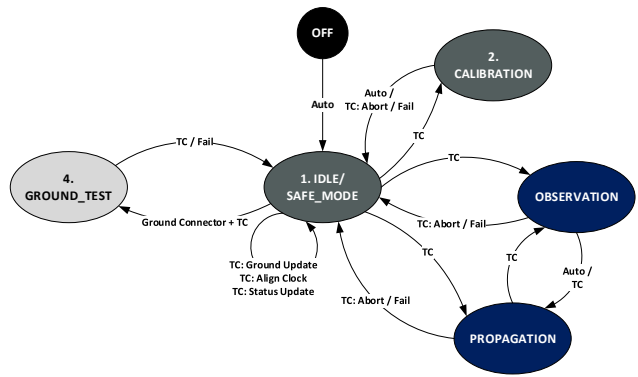

Fig. 1. Operating Modes block diagram

*4.1.1. User Operation and Autonomy*

This section is aimed at defining the preliminary operational use of the unit and discuss its degree of autonomy.

While the desired operation for PODIUM might be similar to the high autonomy modern day star trackers, some fundamental differences need to be taken into account, that will limit the function of the unit. Namely:

- Pulsars are quite uniformly distributed in the plane of the sky, but are not as dense as stars in a star tracker catalogue
- The field of view of PODIUM is necessarily limited. This will imply that in most cases, the telescope won't be pointing at a pulsar a priori
- Observation campaigns require dedicated pointing with good stability over extended periods of time

The consequence of the above items is that the PODIUM measurement acquisition will not be in parallel to the normal operation of a S/C, but will require dedicated slews and pointing modes for the host S/C.

The start of an observation campaign will require the target pulsar to be in the telescope FOV in a stable fashion. The selection of the target pulsar could be carried out internally within the unit, based on the closest observable pulsar of the onboard catalogue, or on the identification of the direction that would benefit most on an improvement of position knowledge. This high-autonomy behaviour would in any case require the unit to provide to the S/C the direction of the pulsar to be observed; the unit would, in a way, be commanding the S/C pointing. Since S/C pointing will answer to operational needs that go beyond the criteria that could be established at PODIUM level, this kind of operation is discarded.

A less autonomous but more operationally feasible approach is favoured, where the observed pulsar ID will be provided to PODIUM as input, the start of observation mode will be triggered by S/C command, and the acquisition of stable observation attitude will be flagged to PODIUM from the S/C. This flag will allow the unit to start acquiring and storing photon times for the generation of phase measurements.

The duration of a measurement campaign will affect the accuracy of the achieved solution. In line with the operational needs of the S/C having priority, the user S/C is assumed to rule the duration of the observation campaign. Once that an observation phase (for a single pulsar) is started, PODIUM will continuously acquire photons. The longer a pulsar is observed, the higher the accuracy of the measurement; still, it must be taken into account that the distance measurement will be referred to a specific time, which at the moment is selected as the observation starting time. In the interval between measurement availability time and measurement reference time will generate a knowledge propagation error when the measurement is included in the Orbit Determination filter; increasing too much the time for photon acquisition will increase the accuracy of the measurement at the reference time but will decrease the accuracy of the information provided at the time at which the measurement is used.

Based on the above discussion, two operation modes are assumed:

- Pull mode: the user S/C will request generation of measurement based on the photon phase data acquired from the start of observation
- Push mode: PODIUM will generate a measurement at a given frequency (for example a measurement every second)

The generated measurement is a distance to the Solar System Barycenter that is then used in the Orbit Determination filter to improve the complete state estimation. The measurement generation and measurement update in the Orbit Determination filter are meant to be asynchronous processes; the orbit propagation part of the Orbit Determination filter is synchronous and providing continuously state solutions to the S/C.

Based on this discussion, Fig. 2 shows the operational diagram foreseen for the use of the PODIUM unit, focusing on the main measurement/functional modes.

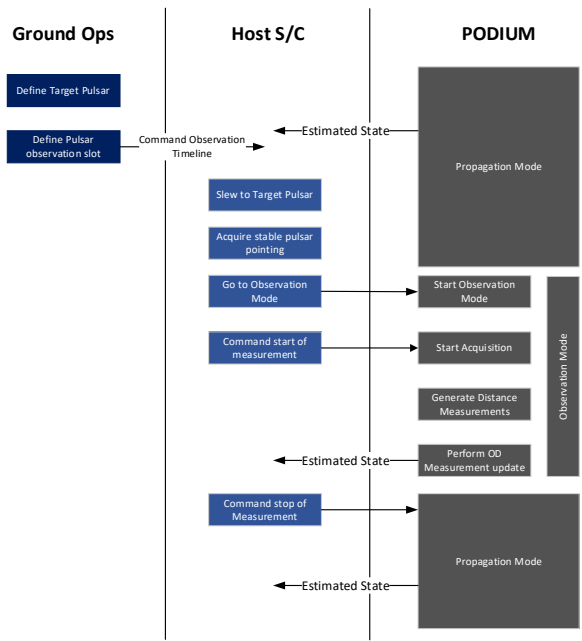

Fig. 2. PODIUM Operational diagram

### 4.2. *Measurement Algorithm Design* and Simulation

The PODIUM activity includes the development of preliminary measurement algorithm and the simulation of the unit performance. Such simulation is achieved via two simulators used in sequence: the first simulator is used to simulate the pulsar pulse reception and the ranging measurements generation, while the second is an Orbit Determination simulator using a performance model of the PODIUM pulsar measurements.

The pulsar measurement simulator, described in this section, is in turn divided in two parts: the photon sequence generation and the pulse processing and measurement generation. The purpose of this section is to present the preliminary breakdown of operations and their implementation, and to support the assessment of the associated computational load.

#### 4.2.1. Photon Sequence Generation

The time at which the photons hit the sensor at the S/C must be simulated to have a realistic scenario for the rest of the computations to be performed over this data.

To do this, the following steps are done:

- Retrieve pulsar data from an observatory: For the example simulation used to describe the algorithms, timing files for the J0030+0451 pulsar have been retrieved from the pulsar catalogue (section 5). These files give the light curve in terms of counts per phase bin, as illustrated with blue circles in Fig. 3.

- Move phase of data from the reference time to the observation time. The phase of the profile is set to the reference time given by the pulsar catalogue for each pulsar. This phase is moved to the observation time using the speed of the light and the difference between the observation and the reference time.

- Perform a multiple gaussian fitting to this data: In the case of the example, two gaussian distributions have been fit to the data using the least squares method (by minimizing the least squares cost function). The number of gaussian distributions to fit to each pulsar is selected manually according to the shape of the pulse (more complex shapes require more gaussian distributions to accurately represent the pulse). The result is illustrated in Fig. 3. This preliminary fitting is part of the onboard Pulsar Catalogue database construction for mission

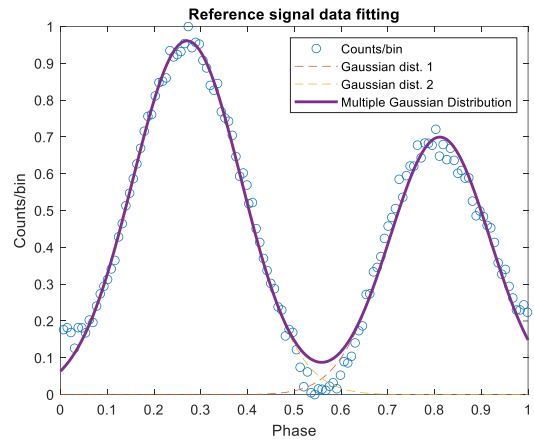

Fig. 3. Reference data multiple Gaussian fitting

- Generate signal photon hits: timestamps are generated from the Gaussian distributions using a Non-Homogeneous Poisson Process (NHPP), at the rates specified for the signal from the pulsar catalogue for the input effective area. These photons are generated at the same reference frame as the original data, usually the Solar System Barycenter (SSB), at the observation time.

- Generate noise photon hits: timestamps are generated using a Homogeneous Poisson Process (HPP) at the rates specified for the noise from the pulsar catalogue for the input effective area. The sum of the signal and noise photons are received by the telescope.

- Transfer photons to the spacecraft: the timestamps are transferred from the SSB to the position of the spacecraft using the light-time between the two points.

- Add clock and detector errors:

o Detector errors: two factors are considered when adding the detector errors. The dead-time and the delay. The delay is simply added to the timestamps. The dead-time is the time that the detector needs to recover from a photon read and to be able to read another photon. It is implemented by checking the difference between consecutive timestamps. If a photon is received within the dead-time span from a previous hit, it is discarded.

o Clock errors: a model of the clock is simulated using Simulink and the characteristics of the selected clock. The error extracted from this simulation is added to the timestamps. A more detailed overview of the clock is presented in section 4.2.2.

*4.2.2. Clock Model*

The clock model simulates the error created by the clock. It is a timing error that will affect directly the precision of the measurements at a rate of 300 m/µs. Therefore, it is of great importance selecting a clock with negligible errors over the time span of an observation. The clock error is built combining three components:

- Clock biases: constant errors over the time of an observation. Affect all timestamps in the same way and result in a distance measurement bias. Can be corrected from ground.
- Clock drifts: not constant during the time of an observation. It is caused by instabilities of the clock and result in a deformation of the profile.
- Clock noise: high frequency error that affects every photon in a different way and results in a noisy signal.

The long-term stability effects can be corrected from the navigation algorithm by observing at least four pulsars, but with a limited accuracy, as this number of observations will take longer time spans than the stability of the long-term effects.

The clock model used (extracted from conversations with Rakon) accounts for these effects and is driven by the following equations:

$$\dot{x}(t) = w(t) + \xi_1(t) \quad (1)$$

$$\dot{w}(t) = \xi_2(t) \quad (2)$$

Where $\xi_1(t)$ is the White Frequency Noise (WFN) and $\xi_2(t)$ is the Random Walk Frequency Noise (RWFN). q1 is the variance of the WFN and q2 the variance of the RWFN and are extracted from the datasheet of the clocks.

The output of the clock simulation for the Rakon RK407 [17], depicted in Fig. 4, represents the expected 3-sigma of the error in seconds (red and orange curves), and the output of the actual simulation (purple), which is used by the simulator to add the error to the timestamps received by the detector.

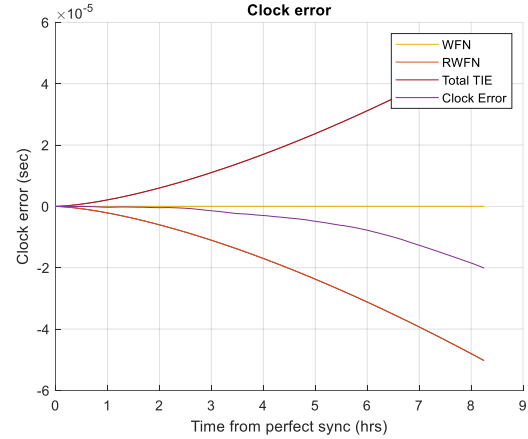

Fig. 4. Clock error from simulation

*4.2.3. Pulse Processing & Measurement Generation*

The steps carried out to simulate the pulse processing to estimate the distance to SSB in the pulsar direction are the following:

- Time-folding of the timestamps: fold all the timestamps into a single period.
- Convert Time of Arrivals (TOAs) into Phase of Arrivals (POAs), using the following expression [9]:

$$\phi(t) = \phi_0 + f(t_0)(t - t_0) + \frac{\dot{f}(t_0)}{2}(t - t_0)^2 + \frac{\ddot{f}(t_0)}{6} + (t - t_0)^3 \quad (3)$$

- Divide counts in phase bins: The previous counts are now divided into phase bins. The size of the bins is defined according to the number of photons received. More photons received mean a clearer signal, allowing smaller bin sizes (more bins) for a more accurate representation of the data. If the number of signal photons received is higher than 1000, the period is divided in 64 phase bins. If the number of signal photons received is between 100 and 1000, the period is divided in 32 phase bins. For less photons, the period is divided in 16 phase bins. These numbers have been set performing a trade-off where different values were tested and can be changed in the main script of the simulator. In this simulation, 8 hours of observation with an effective area of 50 cm2 are simulated. The Counts/bin data can be observed in Fig. 5.
- Fit the signal to the data: Two different methods have been implemented to perform this step. The first consists of performing a multiple Gaussian fitting to the received photons, similarly to the first step performed over the reference data. The second method consists of getting the equation of the fitting performed over the reference data, and simply adjust the phase to the received photons, using the least squares method. It has been found that the second method leads to more accurate results and is the method used by default in the simulator, which has been used to generate the results presented in section 6.1. The outcome of this fitting can be observed in Fig. 5.

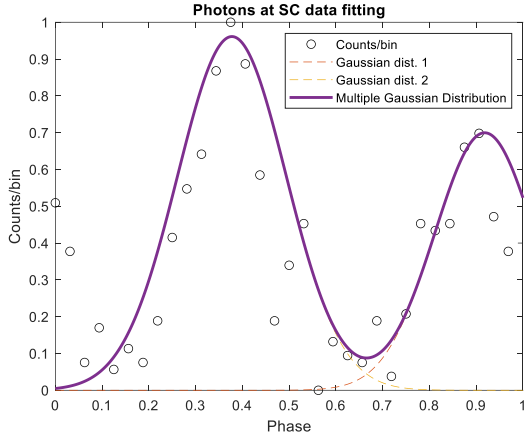

Fig. 5. Received data signal fitting. 8 hours observation. Aeff = 50 cm2. Pulsar: J0030+0451

- Calculate phase shift: The phase shift between the reference signal (moved to the estimated position of the spacecraft) and the signal received by the spacecraft is calculated by comparing the maximums of both signals. The phase shift is illustrated in Fig. 6.

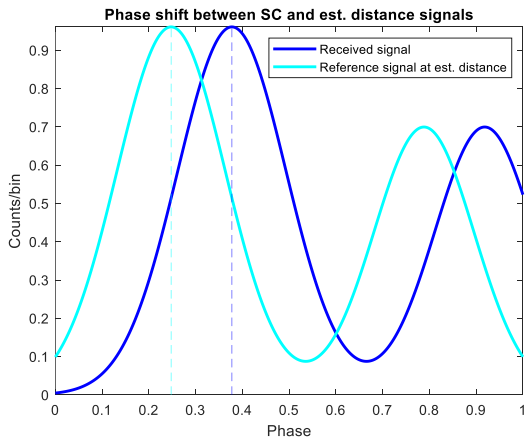

Fig. 6. Phase shift between reference signal at estimated position and actual signal at spacecraft

- Calculate the distance: Using an estimate of the distance within half the distance the light travels during a period, and the phase shift, the distance from the spacecraft to the SSB in the pulsar direction is calculated (the result is simply the sum of the estimated distance and the distance extracted from the phase shift between the two signals):

$$Distance\ (SSB - SC) = \\ Distance\ (SSB - SC)\ est + \\ phase\ shift\ (est - rec) * period * c \quad (4)$$

*4.2.4. Pulsar Campaign Simulator*

To assess the performance of the system with confidence, a campaign of several simulations has been executed for each of the catalogue pulsars.

A batch of 300 simulations have been performed over all the pulsars in the catalogue for 2, 4, 6, 8 and 10 hours of observation time, and for 25, 50 and 80 cm2 of effective area. In the case of the Crab pulsar, 0.25, 0.5 and 0.75 hours of observation time were simulated for the same effective areas, considering its high photon rate.

The errors of each simulation are used to compute a measure of the 3-sigma confidence on the system.

The results obtained with this simulator can be found in section 6.1.

*4.3. Orbit Determination Algorithm Design*

The orbit determination functionality is implemented by an estimation algorithm based on LOTNAV's Square Root Information Filter. More details on the main aspects of the algorithm are provided below and in [10].

*4.3.1. Trajectory Propagation*

The propagation of the trajectory is performed with a RK78 numerical integrator. The propagation of the spacecraft state can be done in any of the modelling worlds of interest for trajectory navigation purposes. Those are:

- **Nominal World**: which emulates a deterministic dynamical system with known parameters. This would be the same as considering a perfect world with perfectly known dynamics and parameters defining them. Propagation with nominal conditions is typically performed for mission analysis tasks.

- **Estimated World**: which emulates a deterministic dynamical system with estimated deviations to the known parameters. The estimated world is the environment of the navigation filter.

- **Real World**: which emulates a dynamical system with stochastic components added to the parameters defining the dynamic interactions, thus allowing the simulation of a system that can be closer to reality. Errors and mismodelling of the nominal world are included here, providing a dispersed trajectory around the nominal one. The estimated world, thus navigation, tries to determine the real world using the observables and modelling the dynamics.

Different dynamic effects are modelled in LOTNAV, such as:

- Gravitational forces: main central body and third bodies, including non-spherical perturbations
- Low-thrust propulsion forces (if required)
- Solar Radiation pressure forces
- Aerodynamic drag forces close to bodies with atmosphere
- Spacecraft residual forces to account for mismodelled effects

More details on the modelling of the dynamics can be found in [10].

*4.3.2. Pulsar Measurement Model*

A measurement model has been implemented to include pulsar-based measurements. For each observed pulsar, the measurement model provides the range between S/C and SSB projected in the direction of the pulsar by exploiting the following equation:

$$c \cdot (t_{SSB} - t_{S/C}) = f(r_{S/C}, \check{n}, D_{PULSAR}, V_{PULSAR}) \quad (5)$$

According to which the delta elapsed time that takes the light to reach the SSB and the spacecraft is a function of the spacecraft position and the pulsar direction. Additional parameters $D_{PULSAR}$ and $V_{PULSAR}$ are functions of the pulsar intrinsic properties which can be neglected in a first approximation. Therefore, navigation corrects the S/C position along the radial direction observed from the pulsar. Using several sources in the navigation campaign considerably improves the knowledge along all directions.

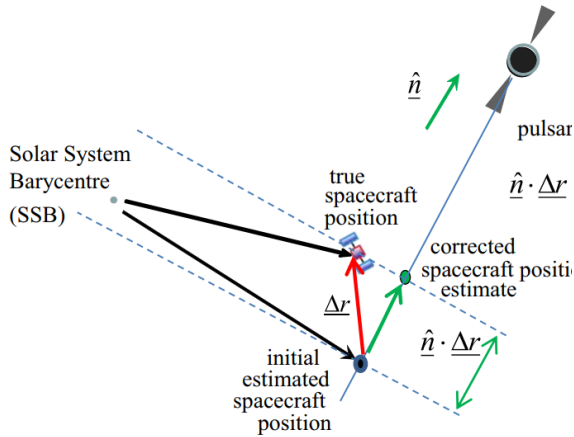

Fig. 7. Graphical representation of a pulsar-based range measurement (credits: [7])

According to the presented model, the implemented equivalent pulsar measurement model performs the following steps:

- Retrieves the current spacecraft state
- Retrieves the ephemerides of the Solar System Barycenter (SSB) and the spacecraft state relative to the SSB.
- Computes the unit pulsar direction and projects the state relative to the SSB in the pulsar direction.
- Perturbs the range measurement according to the propagation world.
- Adds clock error contribution.
- Outputs the estimated measurement.

The clock is modelled according to the following expression:

$$clock\_error = b + q1\sqrt{\Delta T} + q2\sqrt{\frac{\Delta T^3}{3}} \quad (6)$$

Where $b$ is the clock bias and $q1$, $q2$ are the coefficients that describe the time evolution of the error, and their values can be found in the datasheet of the corresponding clock.

*4.4. Physical Design*

This section is devoted to the presentation of the optical, mechanical and thermal design for the PODIUM telescope. The main requirement driving the configuration is the need of fitting the target envelope of 600 x 150 x 240 mm³ and the maximization of the effective area to maximize the accuracy of the measurement over a fixed acquisition time.

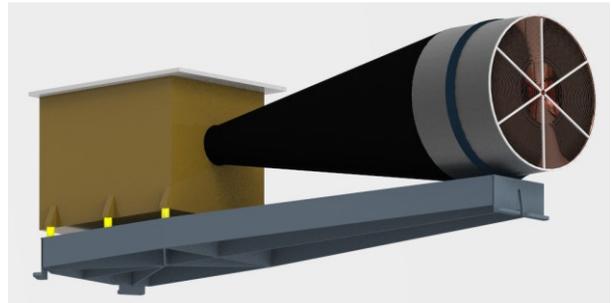

Fig. 8. PODIUM physical configuration concept

*4.4.1. FOV sizing*

The AOCS accuracies for a host system using PODIUM are prescribed in as APE 0.2 deg and RPE 0.05 over 1000s, with 99% probability and 90% confidence interval.

This requirement allows a preliminary sizing of the FOV of the detector. The FOV, in fact, shall be large enough to ensure that the targeted pulsar is within the actual FOV considering AOCS pointing APE and RPE, guaranteeing that no photons are lost due to mis-pointing.

In line with this approach, the FOV size (obtained from the optics/collimator and detector combination) can be proposed preliminarily in the interval between 0.25 deg and 0.5 deg maximum.

A wider FOV size will imply increasing the area exposed to background noise; this will reduce the S/N ratio of the reconstructed pulsar pulse, implying that longer exposure times might be required to achieve sufficient accuracy especially in case of faint sources.

*4.4.2. Telescope Configuration*

The baseline design for PODIUM's telescope is based on Type Walter I design, in which mirrors are composed by a set of nested mirrors in concentric fashion.

The telescope FoV size drives the aperture of the optics and be related to the maximum number of nested mirrors to accommodate. The current target FOV is 0.25 deg aperture.

Typically, combinations of parabolic and hyperboloidal mirrors are used to achieve reflection of collimated rays to a specific focus.

For PODIUM the main function of the optical and detection assembly is to collect and time pulsar photons, with a baseline design that does not require imaging capability (the selected detector is a single-pixel SDD with 9.44 mm diameter, see §4.5.3). Therefore, it is not necessary to focus the light to preserve the incoming light direction.

Not needing focusing it is possible to substitute the parabolic and hyperboloidal surfaces for conical surfaces.

Conical mirrors are easier to manufacture and potentially allow the inclusion of more than two mirror stages in series to reduce the grazing angle at each mirror. Nevertheless, only two consecutive cones per shell are required for PODIUM.

Each shell is defined by the following data

- The longitudinal length L. It is equal for all the cones in the same shell
- The entrance radio R
- The grazing angle θgrazing at the first (entrance) cone. The grazing angle defines the angles for all cones at the same shell because, for a given grazing angle at the first cone, the optimum performances are when the semi-angles of the cones θ1 and θ2 are:
  - θ1 = θgrazing
  - θ2 = 3 x θgrazing

The mirrors are encapsulated in an aluminium cylinder with external diameter 150 mm.

Three design/optimization iterations were carried out for the telescope optical design, resulting in a total of 37 shells distributed in 16 shells with 2 Conical Mirrors in series and 21 shells with a unique Conical Mirror.

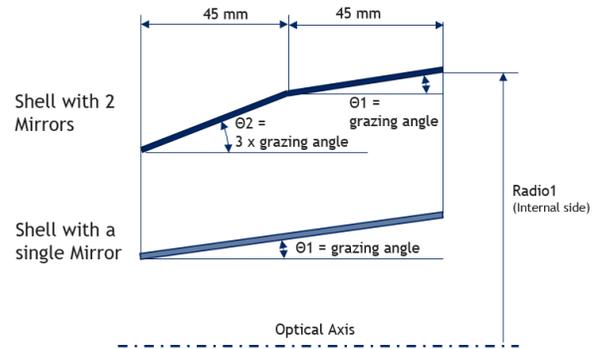

Fig. 9. Shell concept for 2 and 1 cones

The mirrors are encapsulated in an aluminium cylinder with external diameter 150 mm.

Three design/optimization iterations were carried out for the telescope optical design, resulting in a total of 37 shells distributed in 16 shells with 2 Conical Mirrors in series and 21 shells with a unique Conical Mirror.

*4.4.3. Pulsar Energy Range*

PODIUM is initially designed to detect pulsars in the range of energy between 0.5 keV to 10.0 keV. It should be noticed that the larger is the photons' energy, the smaller shall be the grazing angle on the telescope mirrors to have an efficient reflectivity, leading to a more complex telescope design with a large number of shells and lower overall efficiency.

Fortunately, further analysis of the most feasible pulsars sources allowed to establish that the predominant number of pulsars are in a lower range between 0.5 keV to 2.0 keV. In this range, the grazing angle is not so critical, and a simpler mirror system can be designed.

*4.4.4. Mirror Surface*

The mirrors' reflective surface is made of Au and NiCo over a substrate of Al2O3 or Ni:

- Layer 1       Au              25 nm
- Layer 2       NiCo            20 μm
- Substrate     Al2O3 or Ni     200 μm

The next plot shows the reflectivity of the mirrors (ordinate axis) with respect to the pulsar energy in keV (x-axis). The coloured lines represent the grazing angle in arcdeg.

The reflection is acceptable for energy smaller than 2 keV and grazing angles smaller than 1.25 arcdeg. For energies larger than 2.0 keV, the reflectivity decreases significantly.

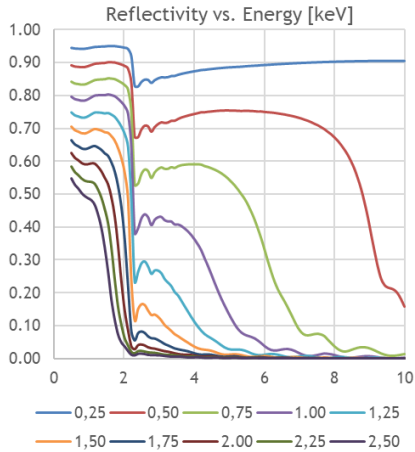

Fig. 10. Reflectivity for the Selected Coating

*4.4.5. Evaluation of the Effective Area*

The effective area depends on several parameters:

- The grazing angle that is constant along all the surface of a shell due to the conical shape.
- The number of cones in series. For each reflection, the energy is reduced.
- The annular surface of each shell
- The energy of the pulsar.

Each pulsar source has different distribution of energy, and this distribution has been considered by weighting the reflectivity for a given energy with the probability to have a pulsar with this specific range of energy.

The result is a unique equivalent effective energy for each of the selected sources. These values are indicated in the table below.

A reference effective area of approximately 60 cm2 is considered for all selected pulsars.

*4.4.6. Focal Plane Design*

The Focal Plane is based on the detector FastSDD-70 from Amptek (see §4.5.3), having a detection surface of 70 mm2 equivalent to 9.44 mm in a unique pixel.

Thanks to the defocusing of the incoming image allowed by the single pixel detector, the Focal Plane alignment tolerances are not critical neither in position nor in focus nor in tilt. The alignment of the Focal Plane can be performed by dimensional measurements instead of optical and by shimming.

The Detector and the Amplifier inside its case conform an element that is attached to an aluminium Plate. The connection of both elements, the Detector and the Supporting Plate, has good thermal conductivity to dissipate the heat of the Detector.

The Supporting Plate is attached to the Optical Bench structure with thermal washers for thermal insulation while there is another thermal link to the Radiator. It should be noticed that the relaxed stability requirements of the Focal Plane allow the use of thermal washers based on fiberglass material.

The Detector is inside an aluminium box that performs two functions. The protection against radiation and it creates a homogeneous thermal environment around the Detector.

Table 2 PODIUM effective area

| PULSAR NAME | EQUIVALENT EFFECTIVE AREA [cm2] |
|---|---|
| B0531+21 | 60,4 |
| B1509-58 | 59,3 |
| B1821-24A | 60,7 |
| B1937+21 | 59,1 |
| J0030+0451 | 61,6 |
| J0218+4232 | 60,9 |
| J0437-4715 | 61,2 |
| J0740+6620 | 62,1 |
| J0751+1807 | 61,1 |
| J1012+5307 | 61,3 |
| J1024-0719 | 61,3 |
| J1231-1411 | 61,8 |
| J2124-3358 | 61,2 |

*4.4.7. Thermo-Mechanical Design*

The design is driven for the size of the main components that are the Telescope, the Detector with the Amplifier, and the Electronics.

The Radiator is located as baseline on top of the Instrument. It is assumed that this position offers the maximum visibility of deep space. It is assumed that the instrument shall thus be mounted on the surface of the host S/C, allowing exposure of the radiator.

The Electronics box is used for shielding the Detector in the lateral sides and is attached to the Radiator directly for thermal conductivity optimization. The Radiator has a thermal link to connect the Focal Plane. It is flexible to minimize the disturbances generated on the Detector.

The Electronics and the Detector are insulated by thermal washers with respect to the Optical Bench and supporting structure.

Supports of the Optical Bench are integrated in the same block than the OB to save mass and integration time. The supports provide large flexibility for differential extension movement thanks to a large plate with low thickness.

A preliminary Finite Element Model (FEM) analysis has been carried out to assess the structural response of the design, and to assess its thermo-mechanical behaviour. The FEM analysis has been used to iterate the design and ensure that the first natural frequency of the instrument is above 60Hz.

The mass budget of PODIUM obtained from the FEM model is presented in Table 3. The Optical Bench is assumed to be made of Aluminium, while the telescope body is in CFRP.

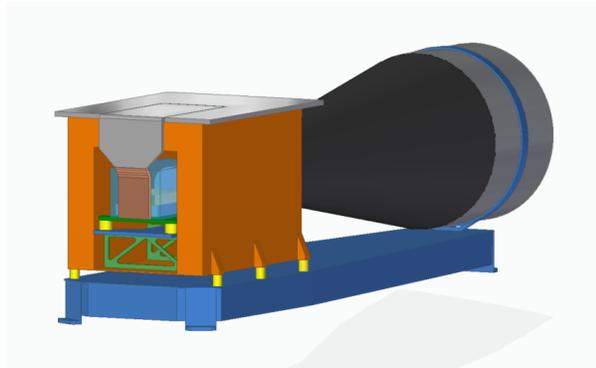

Fig. 11. PODIUM Instrument concept

Table 3. PODIUM mass budget

| Item | Mass [gr] |
|---|---|
| Telescope | 1557 |
| Structural elements | 881 |
| Focal plane | 370 |
| Thermal Control | 661 |
| Electronic Box | 2167 |
| Miscellaneous | 315 |
| **Total** | **5951** |

The thermal design main focus is to ensure dissipation of heat of the Electronics and Focal Plane to keep the temperature inside the operational and survival ranges, depending on the functional mode.

The electronics (detector included) is the power dissipating element, with a maximum power of 12 W.

A radiator is directly attached to the electronics box and to the detector Cover. The detector will have its own internal cooling device, and the radiator is mainly used to dissipate the power of the electronics. The radiator is made of Silver Teflon 10 MIL.

A high conductivity aluminium Focal Plane case is used to create a homogeneous thermal environment at the Detector and Amplifier. The same material is used for the Optical Bench.

Heaters with 5 W installed power are located at the Focal Plane case, close to the radiator, and are used to regulate the temperature of the Electronics when radiation is very high. The purpose of the heaters is to ensure that the temperature of the electronics will be inside the operating range. It is not desired to operate with a very low average temperature of the instrument to minimize power transmission to the platform, that should be limited. The baseline is to have the heaters controlled by the electronic of the instrument to be autonomous.

Multi-Layer Isolation (MLI) covers most external surface, while Single Layer Isolation (SLI) of VDA covers the aperture of the Telescope.

PODIUM is attached to the S/C with FR4 washers for isolation. The final thermal concept is shown in Fig. 12.

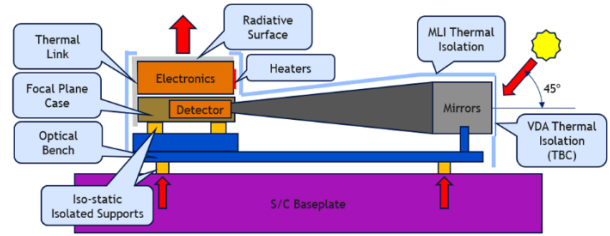

Fig. 12. PODIUM Thermal Concept

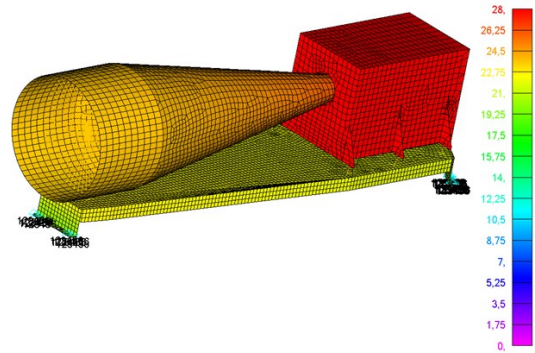

Fig. 13. PODIUM steady state temperature distribution

A thermal analysis was performed, based on the FEM model used for eigen-frequencies determination. The thermal characteristics of the materials have been defined as well as conductivity characteristics of the thermal washers. The FEM model evaluates the temperature distribution for a steady-state case, thus providing the displacement of critical points.

Results show a Line of Sight (LoS) variation of 48 arcsec and a displacement of detector of 600 μm.

The deviation of the LoS of 48 arcsec is not critical because is inside the Field of View of the instrument and much smaller than the pointing accuracy expected for the AOCS system. In addition, as already discussed, the detection of the direction of the pulsar is not relevant.

The LoS deviation is mainly produced by the bending of the Optical Bench due to the load developed due to the differential thermal expansion of the Optical Bench (aluminium) and the telescope cone (CFRP). This load can be reduced adding flexibility in axial direction between the CRFP and the Focal Plane.

*4.5. Avionic Design*

*4.5.1. Software Architecture and Design*

The PODIUM software (SW) will provide the functionality needed by the PODIUM system for:
- Load and start-up of the SW
- Acquisition of data from the detector
- Processing of signals and generation of navigation solution
- Handle telecommands for configuration and send telemetries with its status
- General management and monitorisation

The high-level architecture of the PODIUM software is shown in the following diagram. Fig. 14 shows the flow of information between the components of the SW.

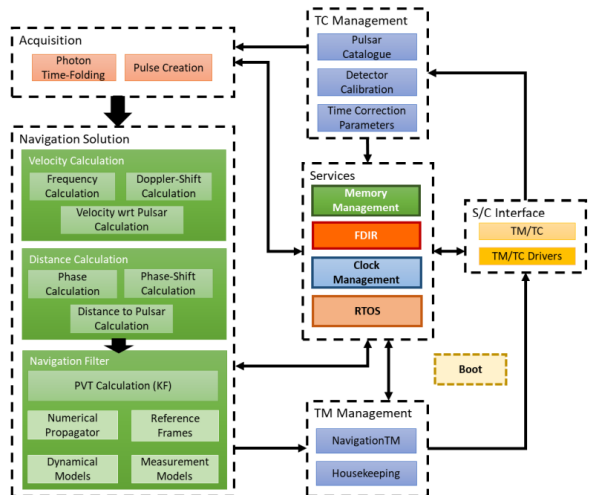

Fig. 14. PODIUM SW Overview

The approach for the SW will be modular. The SW will be composed of several components organised hierarchically by functionality.

The functionality of the software is distributed and decomposed in the following SW components:
- Acquisition: these components prepare the outputs from the detector to provide them to the Navigation Solution. The two functionalities provided by this package are the Photon Time-Folding and the Pulse Creation.
- Navigation Solution: it implements the generation of the absolute position, velocity, and time (PVT). The three main functionalities included are the Distance Calculation, the Velocity Calculation, and the Navigation Filter.
- TC Management: These components handle the telecommands (TCs) received by the SW.
- TM Management: These components manage the telemetries (TMs) sent by the SW to the S/C.
- Services: These components provide different services to the rest of the SW components.
- S/C Interface: Provides the interface with the spacecraft or/and ground.
- Boot: This component provides the functionality for loading the SW and the correct initialization of the rest of the SW components.
- Services: They package groups components that provide different services to the rest of the SW components. These components are memory management, FDIR, clock management, mode management.
- RTOS: The services and drivers needed to run the SW on top of the Real-Time Operating System (RTOS).

*4.5.2. Electronic Design*

The architecture of PODIUM electronic has been selected by considering the following mission requirements/constrains:
- Timing and processing performances
- Low power consumption
- Very compact mechanical configuration due to reduced allocated volume
- Good degree of autonomy for fault management

To achieve the above design goals, the PODIUM electronics design must take advantage by use of complex System On-Chip (SoC) devices and large Field Programmable Gate Arrays (FPGAs) that allow to integrate communication, process, and control functionalities in a small volume.

The PODIUM Electronics encompasses three main functional modules operating without redundancy:
- Power functional module. In charge of receiving power from the S/C platform, conditioning it and distributing internally. From this function the remaining functions, processing and Front End Electronics (FEE) are supplied. FEE in turn will supply/bias the detector in the telescope as needed.
- Processing functional module. In charge of receiving data from the FEE, processing it according to instrument needs and sending the measurements and engineering data to the satellite platform (on-board computer). This function will as well host the high accuracy oscillator to achieve the precise timing references needed by the unit for the photons' timing.
- Front End Electronics functional module. This function will receive the signal from the detector (photons) to be conditioned and transferred to the processing function. The detector supply/biasing function is part of FEE, due to Electromagnetic Compatibility (EMC) reasons it will be accommodated as close as possible to the detector.

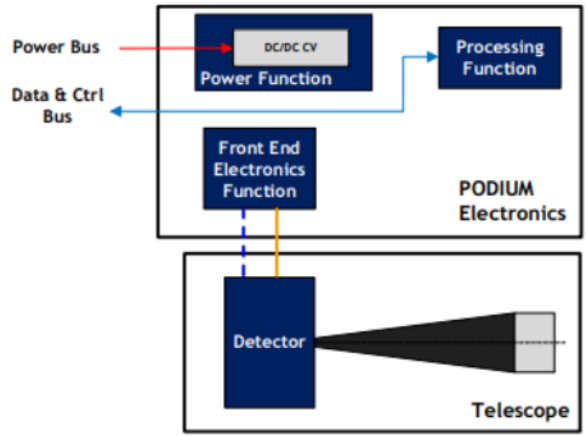

Fig. 15. PODIUM electronics blocks

### 4.5.3. Detector

In order to achieve the high timing accuracy required for PODIUM (better than 1 μs), Silicon Drift Detectors (SDDs) are preferred. The combination of a single-pixel ultrafast SDD detector with the grazing incidence X-ray optics focusing system provides the required effective area.

The FAST SDD® [18] represents Amptek's highest performance silicon drift detector (SDD), capable of count rates over 1,000,000 counts per second (CPS) while maintaining excellent resolution. The FAST SDD® is also available with Amptek Patented C-Series (Si3N4) low energy windows for soft X-ray analysis.

It is the detector used in each of the 56 telescopes that form the NICER instrument [1].

## 5. Pulsar Sky Catalogue

The main factors that must be considered when selecting targets for an X-ray pulsar navigation system are 1) the timing stability of the pulsar (i.e., the predictability of the pulse phase) and 2) the pulsed flux in the X-ray band, which impacts the precision of the pulse times-of-arrival (TOAs) measurements. The PODIUM pulsar catalogue presented here contains 14 objects spread across the sky that were selected based on their rotational stability and brightness in the soft X-rays. Fig. 16 shows their celestial location in Galactic coordinates. Pulsed radio emission is also present in these sources, from which timing parameters have been measured with high precision.

Among the 14 pulsars in the catalogue, 12 of them are nearby millisecond pulsars (MSPs) that have been monitored over the past several years with high-time resolution radio facilities. These observations have been carried out for pulsar timing array experiments such as the European Pulsar Timing Array [11], the North American Nanohertz Observatory for Gravitational Waves [12] and the Parkes Pulsar Timing Array [13].

Together they make up the International Pulsar Timing Array [14] (IPTA), which now includes 65 pulsars. PTAs aim to detect a stochastic gravitational wave (GW) background using an array of high-precision MSPs, and their success relies on the stability of the forming the array and the detectability of the disturbances in the pulse TOAs caused by passing GWs. Sources that have been included in the PTA are the absolute cosmic clocks. Their robust ephemerides can accurately predict the pulsar rotational phase for several years following the last timing measurement. All but one of the 12 MSPs in the PODIUM catalogue are PTA sources. The suitability for inclusion of the non-PTA MSP, PSR J1231-1411, is currently being investigated.

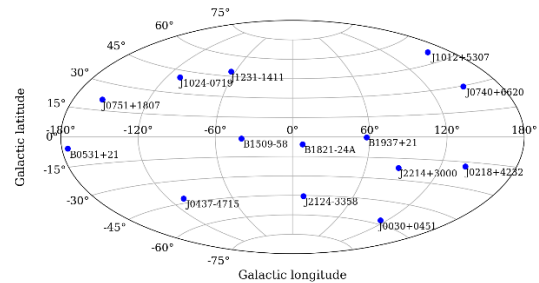

Fig. 16. Galactic distribution of the 14 pulsars contained in the PODIUM catalogue.

The remaining two pulsars in our catalogue are young and highly energetic pulsars: PSRs B0531+21 (the Crab pulsar) and B1509-58. Their high fluxes across the X-ray band and narrow pulse profiles offer the opportunity to measure high-quality TOAs in very short integrations, making them appealing targets for spacecraft navigation. For example, a statistical precision of few tens of microseconds can be achieved for pulsations from the Crab from a 5-min observation with an instrument with capabilities similar to that of NICER. A significant drawback however is their noisy rotations, particularly for the Crab, which limits the predictions accuracy of the extrapolated timing model to only a few days.

### 5.1. Pulsar Parameter Database

Pulsar timing consists of the measurement of the pulse TOAs in high-time-resolution astronomical data and the fitting of these TOAs to a model. The timing model (or ephemeris) relates the measured TOA at the observatory site to the time of emission in the pulsar frame, from which a pulse phase of emission is computed via a model of the intrinsic variations in the pulse period. In addition to the intrinsic rotational parameters of the pulsar, the model includes other deterministic parameters of the source such as astrometric and (if applicable) binary parameters, as well as other information necessary to correct for additional time delays (due to e.g. geometric and light travel-time) and perform general relativistic frame transformations. Other non-deterministic and stochastic components are also included in the model, such as ionospheric, solar system and interstellar plasma dispersion.

Although conceptually straightforward, timing is a highly nuanced technique requiring careful treatment of many fine details. Much of those details have been incorporated in software packages (see [15] and [16]), developed specifically for timing purposes and have been thoroughly tested and validated such that systematic errors are negligible. Thus, once the timing solution of a pulsar has been determined, the model can be used to predict TOAs forward in time. In the absence of stochastic components in the model, the uncertainty associated with the predictions are bounded by the uncertainty on the model parameters.

Pulsar timing ephemerides used in the PODIUM catalogue were obtained from long-term, ground-based radio observations. For the MSPs that are part of PTA programs, we have used the most recent data released, whose high-precision timing models were derived using more than a decade of observations. The timing models of the two young pulsars in the catalogue (Crab and B1509-58) are also based on extensive radio observations carried out for timing purposes. Jodrell Bank has been monitoring the Crab pulsar almost daily since 1984. Lyne et al. (2015) performed a comprehensive analysis of the rotation rate of the Crab pulsar using this large dataset, which showed more than two dozen glitch events and strong timing noise. The result presented in Lyne et al. (2015) is used as the reference timing model for the Crab in the PODIUM catalogue. The second young pulsar, PSR B1509-58, has been monitored at various facilities since 1982, but Parkes began a regular timing campaign on this source in 2007. Results from these observations (Parthasarathy et al. 2020) showed that unlike the Crab, PSR B1509-58 shows much lower levels of timing noise compared to other known young pulsars, and no significant glitch has ever been detected. Thus, the timing model for PSR B1509-58 does not suffer the same rapid degradation in the accuracy of predictions as the Crab.

## 6. Performance Assessment Results
### 6.1. *Measurement* Performance

To assess the performance of PODIUM when observing different pulsars, a simulation campaign has been performed over all the pulsars in the pulsar catalogue. For each pulsar, 300 simulations have been run for different observation durations and different effective areas, but for Crab, where 100 simulations are enough to accurately represent its response due to its high intensity. The selected clock for the simulations has been the Rakon RK407 [17] and the detector the Amptek FAST SDD [18]. The simulation campaign has been repeated for the following effective areas: 25 cm2, 50 cm2, and 80 cm2.

As a result of each campaign the 3-sigma curve of the confidence in the error for each pulsar is extracted.

The evaluation of the performance of the system for the two pulsars that give the most accurate results is presented as an example in Fig. 17 and Fig. 18. The points in the figures represent the error of each of the simulations executed for each of the observation times and effective areas, while the curves represent the 3sigma deviation of the data. It can be seen that although the error is higher (in the case of Crab), the 3sigma value stays low as the error is due to a bias in the fitting of the pulse profiles to the data, and can thus be accounted for when extracting the results.

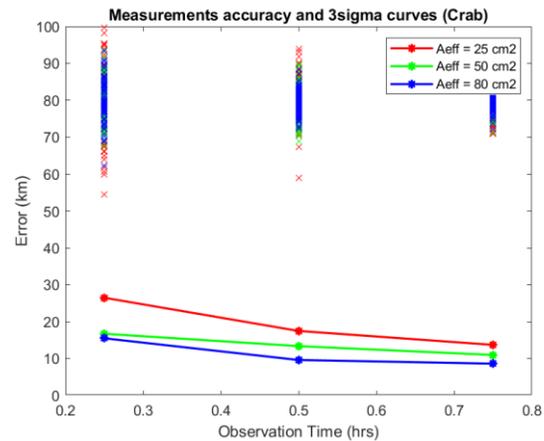

Fig. 17. 3sigma curve for Crab pulsar error

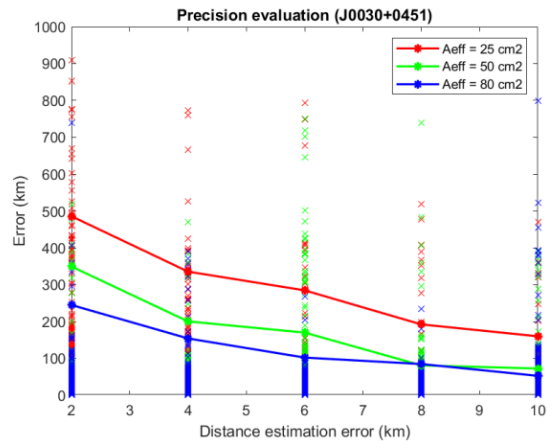

Fig. 18 3sigma curve for J0030+0451 pulsar error

Results show that 6 of the 14 pulsars evaluated are bright enough to provide a significant improvement wrt the initial knowledge error (Crab, B1509-58, J0030+0451, J0437-4715, J2124-3358, J1231-1411).

The rest provide a rate of photons too low to perform good measurements with the effective area of our instrument: the photons received are too scattered for the system to perform a good fitting over them. If the received photons do not represent the original pulse even so slightly, it is not possible to adjust the signal to the received data.

An important aspect of the pulsars is the frequency. The lower it is, the higher the initial knowledge error can be. Certain pulsars with low frequencies (B1509-58, Crab) can be used to narrow the initial knowledge error, so the accuracy of the position can be then improved with measures of higher frequency pulsars.

Frequency also affects the precision of the distance calculation. An error in the phase fitting represents a higher error in the distance in a pulsar with low frequency (a lower frequency means a higher period: an error in the phase means a larger time span if the period is higher. A larger time span of error means a higher distance error).

A special case is the B1509-58 pulsar, which has a very large SNR (SNR = 32.5) and allows to reduce the error from 22720 km to around 200 km. Higher precisions are not achieved due to the low frequency of the pulsar, as previously explained.

The intensity of the pulsar determines along the SNR the observation time needed to accurately reconstruct the pulse. The more intensity, the less time needed.

The shape of the pulses also affects the precision of the system. Diffuse peaks (with high deviations) are difficult to reconstruct as the noise scatter the time-tags of the photons. Very sharp peaks (with low deviations) are difficult to detect as a small number of photons are received from the very moment of the peak.

The non-linearity of the error with respect to the observation time can be attributed to the inclusion of the clock error to the simulations, and to the effect of the errors in the fittings to the data.

The clock error increases with time, not allowing the system to reach negligible errors no matter how high the duration of the observation is. For large observation times, the definition of the signal is higher, but phase shifted, therefore increasing the error.

For pulsars with low count rates (such as J1012+5307 or J1024-0719), and for the effective area of our system (around 50 cm2), the measurements quality does not increase with the observation time, as the received photons are too disperse to reconstruct the original signal. For those pulsars to be useful for a system of this characteristics, very long observations with a very stable clock would be needed.

### 6.2. Orbit Determination Performance
#### 6.2.1. Description of Simulator and Environment

An orbit determination (OD) simulator is used to execute the required performance analysis. The pulsars sky catalogue is used to define the pulsars coordinates, while the outputs of the simulation campaign are used to the define the noise level associated to the measurements of each of the pulsars in the catalogue. The OD algorithm is illustrated in [10].

The simulator is designed with the following architecture:

- LOTNAV is used as an OD simulator, which performs Monte-Carlo analysis given the trajectory data, the scheduling of the measurements and the associated noise level. The OD algorithm used by LOTNAV is illustrated in [10].
- A Python orchestrator has been used in order to perform the following tasks:
  o Wrap the call to LOTNAV, allowing to easily set inputs and retrieve outputs.
  o Define the observations scheduling and retrieve the measurement noises associated to the given input acquisition times.
  o Retrieve the outputs of the MC analysis and plot.

#### 6.2.2. Results of the Orbit Determination Simulation Campaign (Interplanetary Flyby Scenario)

A baseline analysis has been defined with the following configurations:

- Availability = 100% (pulsars signal acquisition can be performed during the whole day)
- Acquisition duration = 8 hours
- Effective instrument area = 80 cm2
- Full pulsars catalogue

This analysis provides a best estimate of the system performance. Then, parameters are varied independently in order to assess the impact each one has on the achievable OD accuracy.

Fig. 19 shows that the results of the different MC runs are not much dispersed and that the achievable position total position knowledge is in the order of 10 km, while the velocity knowledge is maintained below 1 cm/s at steady state. It is also relevant to observe that the visibility of the clock error dynamics is very reduced, due to the size of the position error. The flyby epoch is clearly highlighted in the plot by the peak close to half of the mission time (around 50 days) which raises the uncertainty to very high values which the filter is capable of compensating within a few hours/days.

A first assessment is performed on the impact of the catalogue choice. Besides the Crab pulsar, which provides the best performances by far, there are only a few pulsars which are capable of providing measurement errors below the 10 km level. In order to understand the impact of this, an analysis has been performed by removing two of the three accurate pulsars from the catalogue. The output is shown in Fig. 20, where it is clear that the performance is much worse than the previous case, with an average knowledge level around 100 km and velocity uncertainty up to 10 cm/s.

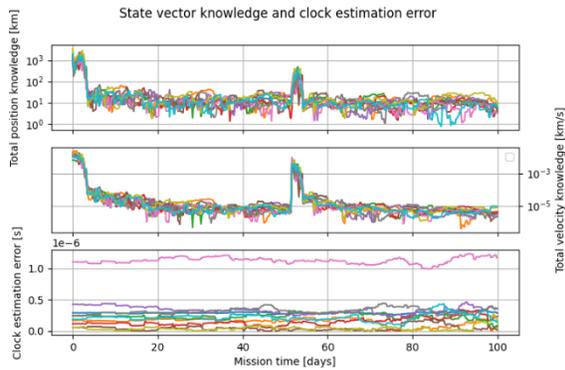

Fig. 19. Flyby - Baseline OD performance. Acquisition time = 8h, full catalogue, availability = 100%, Effective area = 80cm2

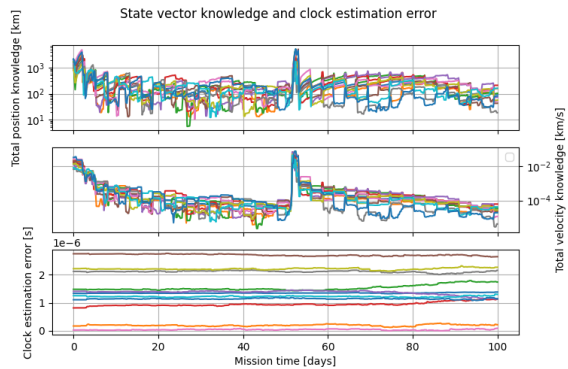

Fig. 20. Flyby - Baseline analysis configuration. 'B0531+21', 'B1821-24A', 'B1937+21' removed from catalogue

The second parameter to be analysed is the effective area of the instrument. Moving from 80 cm2 down to 50 cm2 increases the pulsars position uncertainty. The expected design effective area will lie between these two values thus the following is a worst-case analysis for this parameter. Currently the references effective area is 60 cm2. Fig. 21 shows how the performance is very similar to the one obtained in the baseline analysis (80 cm2 effective area), and this is due to the fact that the error level of the three accurate pulsars in the catalogue increases only slightly while reducing the effective area. Since the analysis is driven by the measurements of such pulsars, the loss of performance is not significant.

As a third analysis, the effect of the system availability is investigated. The navigation system availability for observations is limited in this case to 25%, meaning that a much smaller number of observations can be processed. The output of the analysis, shown in Fig. 22, clearly shows that this parameter has a high effect on the transient phases of the estimation and a much lower effect on the steady-state performance. Indeed, with the much smaller number of measurements available, it takes the filter around 10 days to recover from the flyby uncertainty peak. This outcome clearly states the importance of an effective scheduling of the measurements. If during the interplanetary cruise a small number of observations can be enough, a more intense campaign needs to be planned close to the critical points of the trajectory.

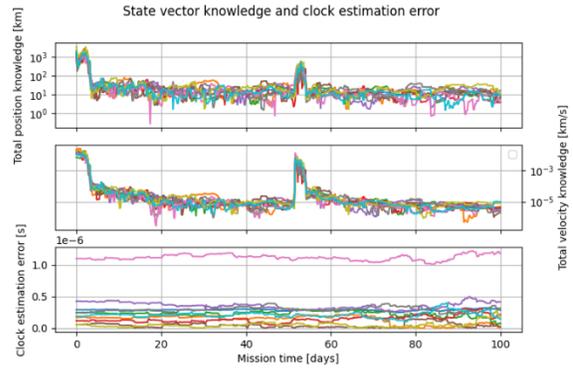

Fig. 21. Flyby - Acquisition time = 8h, full catalogue, availability = 100%, Effective area = 50cm2

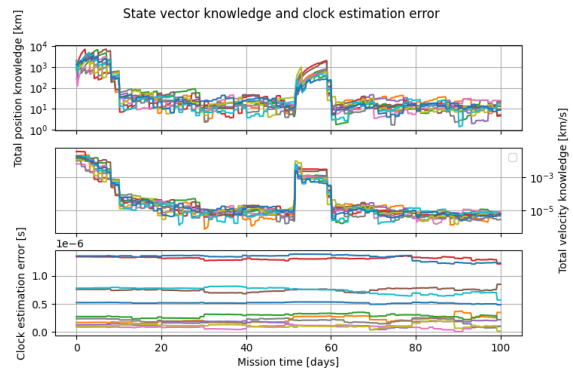

Fig. 22. Flyby - Acquisition time = 8h, full catalogue, availability = 25%, Effective area = 80cm2

The final parameter to be investigated is the acquisition duration. Decreasing the acquisition duration from 8 hours to 4 hours doubles the number of available measurements but reduces their accuracy. By looking at the output of the simulation campaign, it is clear how there is only one pulsar, besides Crab, capable of keeping good performances with effective area of 80 cm2. If the effective area is 50 cm2, only Crab can provide accuracies below 10 km. For the sake of completeness, this analysis has been performed both with effective area of 80 cm2 and 50 cm2. Fig. 23 and Fig. 24 show the outcome of the two simulations and the results show a clear difference. When the effective area is 80 cm2, the presence of two accurate pulsars can keep the average knowledge in the order of 10 km, thanks to the high number of measurements. On the other hand, when only Crab measurements are accurate, the knowledge raises always above 10 km.

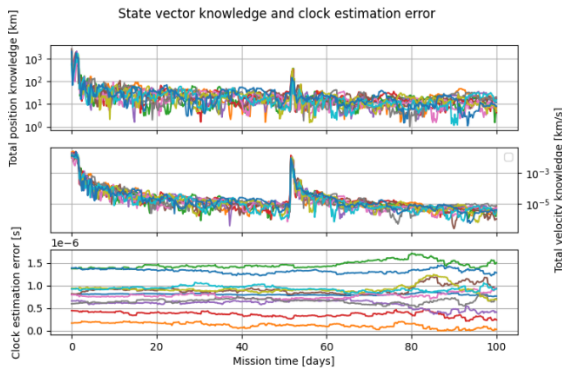

Fig. 23. Flyby - Acquisition time = 4h, full catalogue, availability = 100%, Effective area = 80cm2

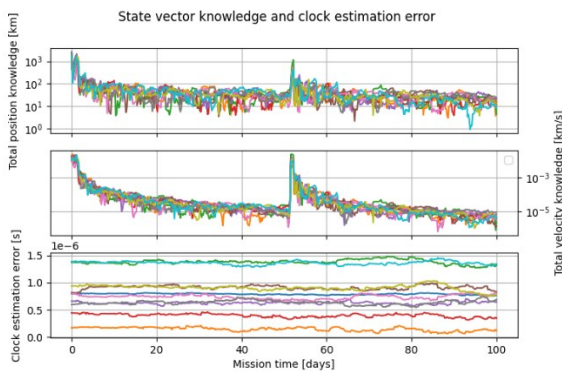

Fig. 24. Flyby - Acquisition time = 4h, full catalogue, availability = 100%, Effective area = 50cm2

*6.2.3. Orbit Determination Simulation Campaign Conclusions*

The system performance has been tested through the OD simulator under different conditions an L2 observatory scenario and an interplanetary flyby scenario. As the results were similar and for the sake of simplicity, only the flyby scenario has been presented. Some of the relevant conclusions that can be drawn are the following:

- The best achievable performance with the current version of the pulsar catalogue is to keep the position knowledge around 10 km and the velocity knowledge below 1 cm/s, which is considered acceptable for routine operations in the mentioned scenarios.
- The inclusion of, at least, a small number of objects with higher accuracy (achieved thanks to the high flux of the observed pulsars, Crab for instance) in the positioning is key to keep error levels closer to the 10 km bound.
- The effective area of the telescope does not have a major impact in the 25cm2 to 80cm2 range, relatively to the performance achievable with the current catalogue version. What drives the accuracy based on the results obtained is the pulsars intensity.
- The availability i.e., the scheduling of the measurements, strongly affects the transient phases of the estimation. An important aspect of such scheduling would be to include extensive acquisition campaigns whenever the a-priori knowledge is expected to be very high. This happens, for instance, at the beginning of the simulations and immediately after the flyby.
- The acquisition time has a visible effect on the performance. Shorter acquisitions have lower accuracy but allow for more measurements to be processed, slightly counteracting the performance decrease. Nevertheless, the achievable performance is clearly worse than the baseline case.

## 7. Conclusions

The work performed on PODIUM successfully produced a feasible preliminary design for an autonomous interplanetary navigation unit.

PODIUM is based on a Wolter Type I telescope with conic mirrors, with a field of view of 0.25 deg and effective area of approximately 60 cm$^2$. A single pixel SDD detector without imaging capabilities is employed. The overall mass of the unit is below 6 kg, with a volume of approximately 600 × 150 × 190 mm. A preliminary measurement generation algorithm has been designed and implemented. The OD performances have been tested with the Pulsar Sky Catalogue produced, and for L2 orbiter and planetary fly-by scenario, with performances in the order of 10 km accuracy when bright pulsars are included in the estimation.

As future activity the development of a breadboard for the pulsar acquisition and measurement generation chain for hardware-in-the-loop demonstration of the functionality.